\newcommand*\patchAmsMathEnvironmentForLineno[1]{%
  \expandafter\let\csname old#1\expandafter\endcsname\csname #1\endcsname
  \expandafter\let\csname oldend#1\expandafter\endcsname\csname end#1\endcsname
  \renewenvironment{#1}%
     {\linenomath\csname old#1\endcsname}%
     {\csname oldend#1\endcsname\endlinenomath}}%
\newcommand*\patchBothAmsMathEnvironmentsForLineno[1]{%
  \patchAmsMathEnvironmentForLineno{#1}%
  \patchAmsMathEnvironmentForLineno{#1*}}%
\AtBeginDocument{%
\patchBothAmsMathEnvironmentsForLineno{equation}%
\patchBothAmsMathEnvironmentsForLineno{align}%
}
\patchAmsMathEnvironmentForLineno{equation}

\documentclass[aps,showpacs,superscriptaddress,floatfix,nofootinbib,twocolumn,fleqn,longbibliography]{revtex4-1}

\usepackage[mathlines]{lineno}
\usepackage{amsmath,graphicx,verbatim,epsfig,amssymb,dsfont}
\usepackage{epstopdf}
\usepackage[latin1]{inputenc}
\usepackage{datetime}
\usepackage[colorlinks=true,linkcolor=blue,citecolor=blue]{hyperref}
\usepackage[usenames,dvipsnames]{color}
\usepackage{setspace}
\usepackage[mathscr]{eucal}

\usepackage{float} 
\usepackage[caption=false]{subfig} 

\newcommand{\be}{\begin{equation}}
\newcommand{\bea}{\begin{eqnarray}}
\newcommand{\eea}{\end{eqnarray}}
\newcommand{\ee}{\end{equation}}

\newcommand{\nn}{\nonumber}

\def\tr{\mathrm{Tr}}

\newcommand{\m}{\mu}

\newcommand{\cd}{\!\cdot\!}

\usepackage[normalem]{ulem} 

\begin{document}

\title{Quantum Simulation of Light-Front Parton Correlators}

\author{M.G. Echevarria}
\email{m.garciae@uah.es}
\affiliation{University of Alcal\'a, Dep. of Physics and Mathematics, 28805 Alcal\'a de Henares (Madrid), Spain}

\author{I.L. Egusquiza}
\email{inigo.egusquiza@ehu.es}
\affiliation{Department of Physics,
University of the Basque Country UPV/EHU, Apartado 644, 48080 Bilbao, Spain}

\author{E. Rico} 
\email{enrique.rico.ortega@gmail.com}
\affiliation{Department of Physical Chemistry, University of the Basque Country UPV/EHU, Apartado 644, 48080 Bilbao, Spain}
\affiliation{IKERBASQUE, Basque Foundation for Science, Plaza Euskadi 5, 48009 Bilbao, Spain}

\author{G. Schnell}
\email{gunar.schnell@desy.de}
\affiliation{Department of Physics,
University of the Basque Country UPV/EHU, Apartado 644, 48080 Bilbao, Spain}
\affiliation{IKERBASQUE, Basque Foundation for Science, Plaza Euskadi 5, 48009 Bilbao, Spain}
\date{\today}


\begin{abstract}
 The physics of high-energy colliders relies on the knowledge of different non-perturbative parton correlators, such as parton distribution functions, that encode the information on universal hadron structure and are thus the main building blocks of any factorization theorem of the underlying process in such collision. These functions are given in terms of gauge-invariant light-front operators, that are non-local in both space and real time, and thus intractable by standard lattice techniques due to the well-known \emph{sign problem}. In this paper, we propose a quantum algorithm to perform a quantum simulation of these type of correlators, and illustrate it by considering a space-time Wilson loop. We discuss the implementation of the quantum algorithm in terms of quantum gates that are accessible within actual quantum technologies such as cold atoms setups, trapped ions or superconducting circuits.
\end{abstract}

\maketitle 

\section{Motivation}

Quantum chromodynamics (QCD), the quantum field theory of the strong interaction between quarks and gluons, has been an incredibly successful but at the same time challenging part of the Standard Model of particle physics. The strong force, mediated by gluons, acts therein on particles that carry one of the three color charges within an SU(3) symmetry group. The success of QCD is for instance manifested in precise predictions of high-energy phenomena based on factorization theorems. These separate the computational description of observables---such as scattering cross sections---into calculable matrix elements on one hand, and on the other, corrections arising from the change of energy or factorization scale of the process (``evolution''), starting from presently often non-calculable but universal quantities, which parameterize, among others, the composition or formation of those hadrons---such as protons---involved in the process~\cite{Collins:2011zzd}. The latter aspect constitutes one of the challenges: a long history of experimental as well as theoretical analyses have revealed a tremendously rich internal structure of the proton. On the other hand, QCD has so far failed to provide an equally reliable tool for precision calculation of a seemingly simple ground state, quite in contrast to the hydrogen atom in the framework of quantum electrodynamics (QED). Part of the challenge is the non-Abelian nature of QCD, with gluons (the gauge bosons) themselves carrying color charges, again quite in contrast to QED. This leads to color interaction not only between quarks, but also between quarks and gluons or even just between gluons, providing a mechanism for peculiar aspects of hadron structure and formation, such as the still hypothetical glue-balls, or confinement.

The modern view of the proton structure goes far beyond the original quark-parton model of collinear moving quarks (and gluons) in a highly energetic proton~\cite{Bjorken:1969ja, Feynman:1969ej}. It now includes correlations between the various spin orientations of the  parent proton as well as of its constituents and the constituents' longitudinal and transverse momentum components (or even position), where the latter are with respect to the \(x^+\) (``light-front time") light-front direction.\footnote{The light-front coordinate system~\cite{Dirac:1949cp}, with \(x^\pm \equiv (x^0 \pm x^3)/\sqrt{2}\) and \(x^\perp \equiv (x^1, x^2)\), where \(x=(ct,\vec{x})\), is a natural choice for describing high-energy interactions~\cite{Ji:1992ku,Wilson:1994fk,Burkardt:1995ct}.} These correlations are typically cast in terms of parton distribution functions (PDFs), or---in the particular case of including transverse-momentum degrees of freedom---transverse-momentum-dependent PDFs. They are complementary to other characteristics such as form factors or generalized parton distributions (see, e.g., Ref.~\cite{EPJA52}). All these functions encode the multi-dimensional structure of nucleons in terms of different correlations between the momentum/spin of the considered parton and its parent hadron, and are currently constrained through experimental data.

A recurring challenge in this respect is the formulation of physical quantities in a quantum field theory. Basically all high-energy probes, hadron-structure studies in deep-inelastic scattering (DIS), or searches for physics beyond the Standard Model in proton-proton collisions at the Large Hadron Collider, involve non-local operators. To complicate issues, these operators are not only separated in space but often involve light-like and thus real-time separation as well. 

\begin{figure*}[ht!]
\begin{center}
\includegraphics[width=0.3\textwidth]{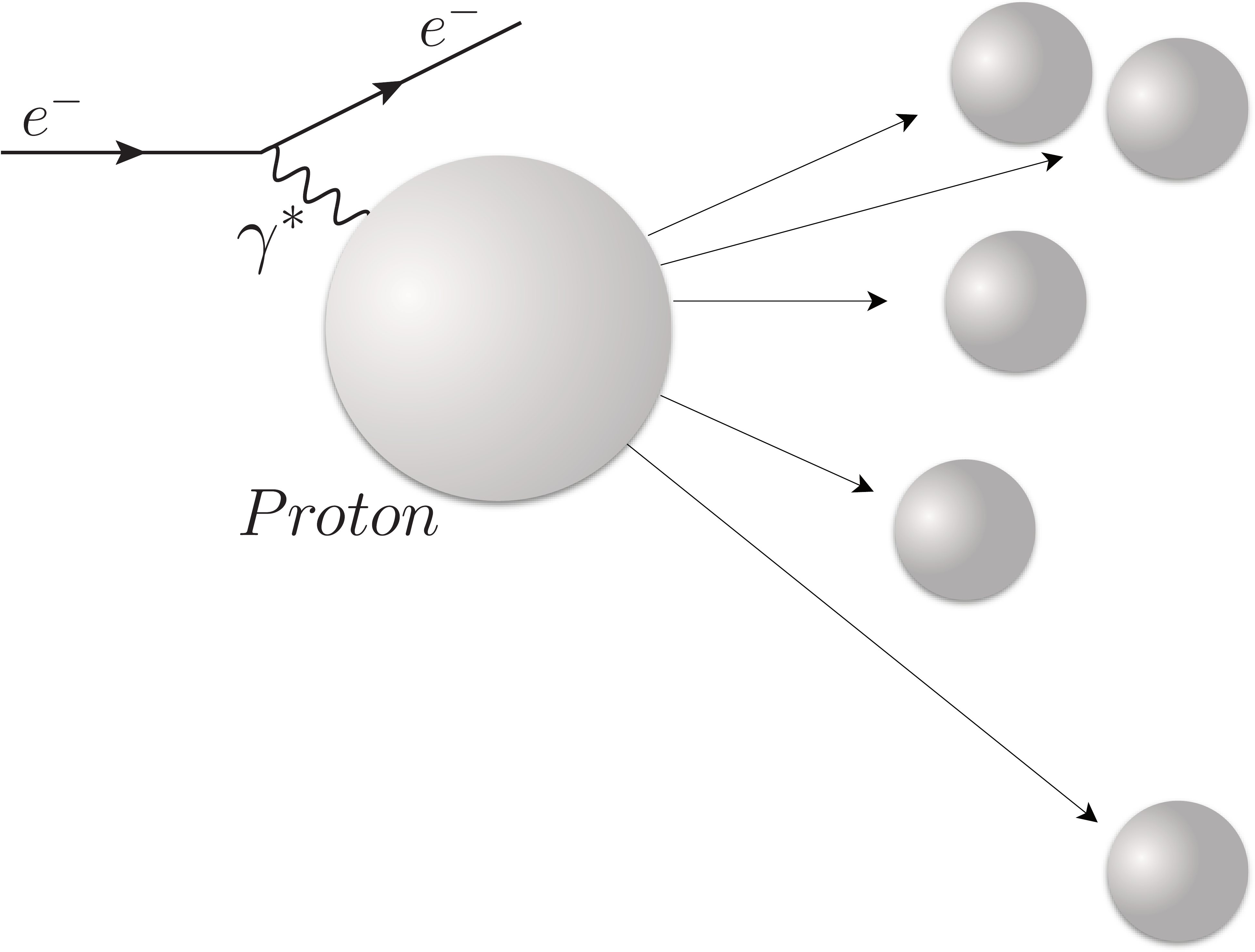}
\qquad
\includegraphics[width=0.3\textwidth]{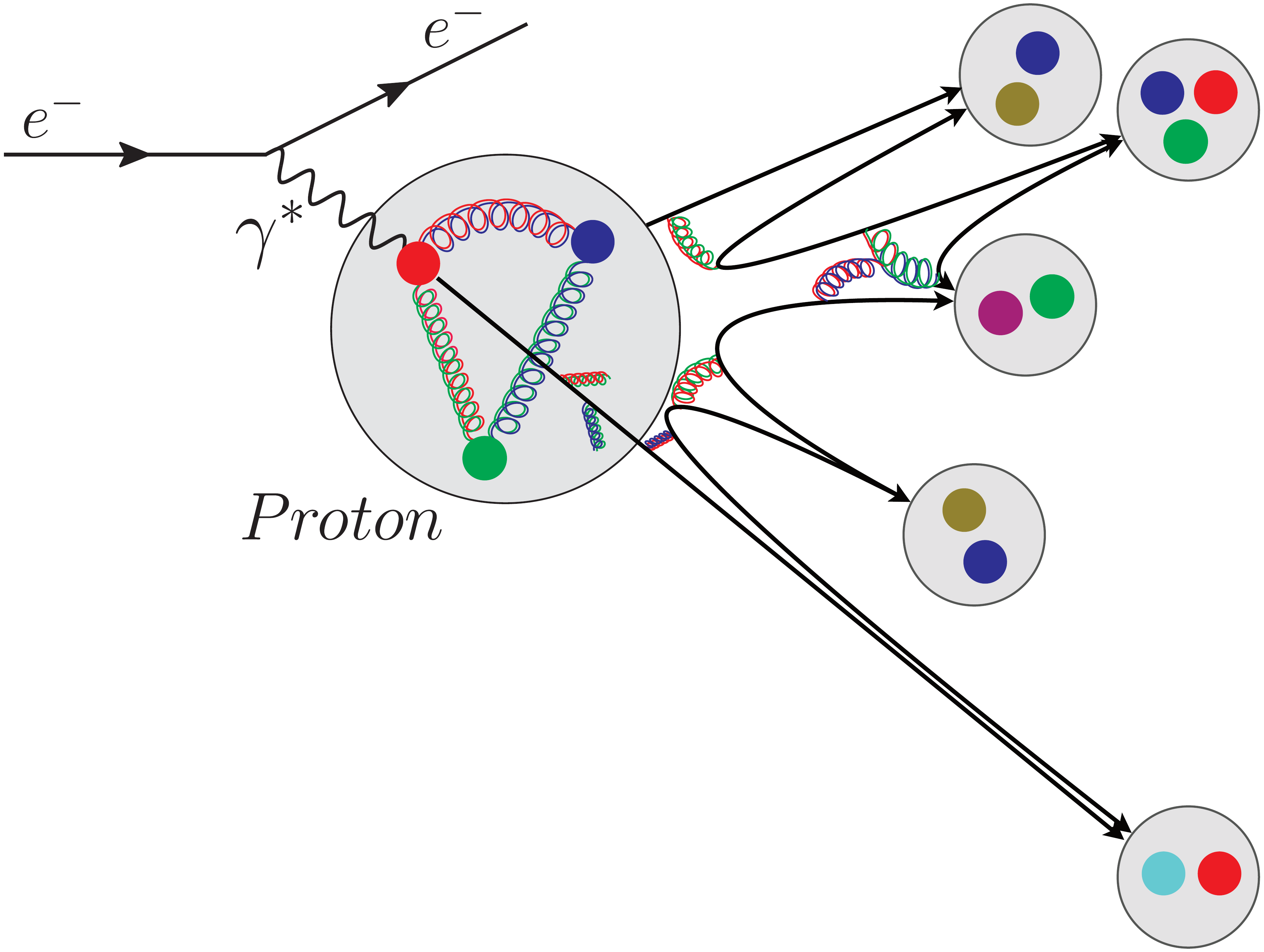}
\qquad
\includegraphics[width=0.3\textwidth]{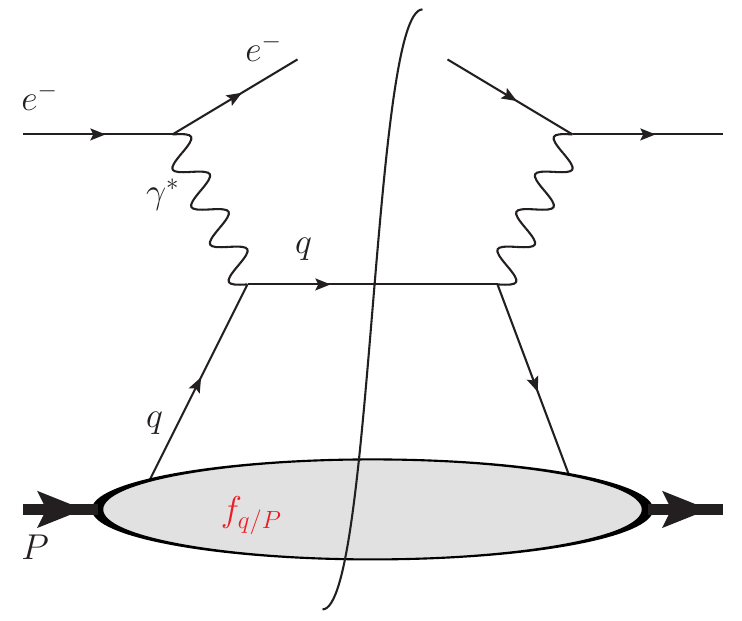}
\caption{Schematic view of the deep-inelastic-scattering process. Left panel: cartoon of the experimentally observable initial- and final-state particles. Middle panel: sketch of the partonic interpretation in terms of interactions between elementary particles. Right panel: its field-theoretical description as a leading-order Feynman diagram, which depicts the factorization in terms of the partonic process, i.e., electron-quark scattering through the exchange of a virtual photon $\gamma^*$, and the non-perturbative parton distribution function, which gives the probability to find such a quark inside the proton (see the text for more details). The vertical cut and the resulting mirror symmetry implies summation over all possible final hadronic states, leaving only a dependence of the cross section on the proton structure.}
\label{fig:DIS}
\end{center}
\end{figure*}

Let us consider the DIS process in more detail.  
It can be proven that in the so-called Bjorken limit, the cross-section \(\sigma\) for DIS (see Fig.~\ref{fig:DIS}) can be approximated as a factorized product of a partonic cross section, \(\hat{\sigma}_f\), which can be calculated perturbatively and describes the elementary scattering of a lepton and an $f$-flavor quark, and the non-pertubative PDFs for an $f$-flavor quark, \(f_{f/P}\), which characterize the partonic structure of the nucleon and gives the probability to find a parton of flavor $f$ inside the proton.
More explicitly, the cross section (where a sum over all parton flavors that can contribute to the process is performed) can be written as (see, e.g., Ref.~\cite{Collins:1989gx}):
\begin{align}
\sigma(\xi,Q^2) &=
\sum_{f}
\int_{\xi}^{1} \! \mathrm{d}\bar\xi\,
\hat\sigma_f(\bar\xi,Q^2)\,
f_{f/P}(\xi/\bar\xi)
\nn\\
&
~~ + {\mathcal O}\bigg(\frac{\Lambda_{\rm QCD}}{Q}\bigg)
\,.
\label{crossec}
\end{align}
Here, $\xi=Q^2/(2pq)$, with $p$ the momentum of the proton, and $-Q^2=q^2$ is the squared invariant mass of the exchanged virtual photon of momentum $q$. 
In order to keep power corrections under control, i.e., for the factorization theorem to be a good approximation, $Q^2$ should be large (larger than the typical infrared QCD scale $\Lambda_{\rm QCD}\sim 1$~GeV).~\footnote{More precisely, the Bjorken limit corresponds to large photon virtuality \(Q^2\) and squared hadronic center-of-mass energy \((p+q)^2\), with \(\xi\) staying fixed.}
The operator definition of the quark PDF, e.g., appearing in the integrand of Eq.~\eqref{crossec}, is
\begin{align}
\label{eq:PDF}
f_{f/P}(\xi) &=
\sum_S
\int\frac{\mathrm{d}y^-}{2\pi} 
e^{-i\xi p^+y^-}
\nn\\
&~~ \times
\langle PS|
\big[\bar \psi ~\mathcal{U}\big](y^-)
\frac{\gamma^+}{2}
\big[\mathcal{U}^\dagger\psi\big](0)|PS\rangle
\,,
\end{align}
where $f_{f/P}(\xi)$ gives the number density of unpolarized quarks of flavor $f$ with a longitudinal fraction $\xi$ of the proton momentum $p$ inside an unpolarized nucleon, which has spin $S$. Here, $y^-$ ($p^+$) is the $-$ ($+$) light-front coordinate (momentum), $\gamma^+$ is the adequate Dirac matrix to single out unpolarized quarks, $|PS\rangle$ denotes the proton state, and $\psi$ the quark field. The Wilson line $\mathcal{U}$ ensures gauge invariance when bracketing wave functions at different space-time coordinates ($0$ and $y^-$, separated here on the light-front). In general, the actual path of the Wilson line depends on the quantity of interest and process used as the probe. In the case of DIS, one has a future-pointing Wilson line~\footnote{A generic vector $a^\mu$ is decomposed as $a^\m=a^+ n_+^\m + a^- n_-^\m + a_\perp^\m$ with $a^+=n_-\cd a$, $a^-=n_+\cd a$, $n_+=(1,0,0,1)/\sqrt{2}$, $n_-=(1,0,0,-1)/\sqrt{2}$, $n_+^2=n_-^2=0$, and $n_+\cd n_-=1$.}
\begin{align}
\mathcal{U}(y) = P\exp\Bigg[
-ig_s\int_{0}^{\infty} \! \mathrm{d}s\,
n_-\cd A(y+sn_-)
\Bigg]
\,,
\end{align}
where $P$ denotes path ordering, $g_s$ the strong coupling, and $A$ is the gauge field. Physically, a Wilson line accounts for an infinite number of gluon emissions from a fast-moving parton, parallel to its direction of motion. For the PDF in Eq.~\eqref{eq:PDF}, relevant for DIS, one can see that the path followed by the Wilson lines consists in a Wilson line that goes from 0 to infinity in the $-$ light-cone direction, and then comes back from infinity to $y^-$. By contrast, for the Drell-Yan process---the annihilation of a quark and an antiquark from two colliding protons into a virtual photon, subsequently decaying into a lepton pair---the path extends to negative infinity and then back to $y^-$. Nevertheless, these two PDFs with seemingly different paths turn out to be exactly the same. They are thus universal, i.e., PDFs constrained in one process can be used directly in the calculation of another process, which makes the formalism used here predictive and so attractive. It may be noted that this strict universality is valid for a selected class of such non-perturbative quantities, while others might be subject to a certain degree of process dependence. 

This analysis can be generalized to other and in parts more complicated processes, such as Higgs-boson or jet(s) production in proton-proton collision, fragmentation into hadrons in DIS, electron-positron annihilation, or proton-proton collision, as well as including dependence on the polarization(s) of involved hadrons and/or partons. In all these processes, if factorization can be proven to hold, the general structure is always schematically given by
\begin{equation*}
\sigma =
\hat\sigma \otimes
\left[\textrm{non-perturbative function(s)}\right],
\end{equation*}
where $\hat\sigma$ is the perturbatively calculable partonic version of the full cross section, the $\otimes$ stands for the necessary kinematical convolutions, and the non-perturbative functions are the relevant ones for the considered process. 
These comprise (un)polarized PDFs, transverse-momentum-dependent functions (TMDs), generalized parton distributions (GPDs), jet functions, fragmentation functions, to name a few. 

Most of the non-perturbative hadronic quantities are given in terms of non-local operators in both space and time. They cannot be calculated in perturbation theory, and two main procedures have been followed to determine them. 

On one hand, global QCD analyses, which make use of factorization theorems, model them and fit the parameters using experimental data. Thanks to their universality, they can be extracted from a given set of processes and applied in others. However, this approach has several limitations, starting from the lack of data to fully constrain their functional dependence, model bias, and limited precision of factorization theorems. 

On the other hand, lattice QCD, which has evolved into a very successful tool during the last decade, in particular for calculating various static properties of the proton. However, the main problem for lattice QCD in treating PDFs, TMDs, GPDs and alike, which are given in terms of non-local operators in space-time, is that it faces the well-known {\em sign problem}~\cite{PhysRevLett.94.170201} which, in principle, prevents Monte Carlo techniques from being applied.

Time dependence in lattice QCD, as needed for dynamic properties, is currently achieved only via detours (see, e.g., the recent reviews \cite{Lin:2017snn,Cichy:2018mum,Constantinou:2020pek,Lin:2020rut}).
One of the traditional and most widely used techniques consists in calculating the Mellin moments of the distributions, but this allows only their partial reconstruction through an operator product expansion (OPE). In addition, it is limited by the practical challenge to reliably calculate higher moments, since the signal-to-noise ratio rapidly decreases and an unavoidable power-law mixing starts beyond the third non-trivial moment. Alternatively, other approaches have been developed in the last years, the so-called pseudo-distributions within the Large Momentum Effective Field Theory (LaMET)~\cite{Ji:2013dva,Ji:2014gla,Ji:2020ect} being the one that has attracted more attention. Within this approach, the light-cone PDFs (and alike) are obtained through an OPE onto their corresponding space-like operators. However, even if promising, this approach is still under development and faces several theoretical and computational challenges, some of them shared with standard lattice calculations, which prevent it from being able to achieve in the near future a reliable calculation of a full PDF (see, e.g., \cite{Schlemmer:2021aij} for a discussion of the different sources of uncertainties and their size in a typical calculation with pseudo-distributions).
Anyhow, all these {\em classical} simulations require vast amounts of computing resources.

Therefore, already during the early times of lattice QCD, the use of quantum simulators and quantum computers to overcome these problems had been put forward. Regarding the newer approaches in lattice QCD, any alternative computational framework that can provide at least benchmarks will also be welcome. But only with the advent of modern quantum technologies does it appear possible to solve problems in QCD where classical approaches fail or face enormous computational requirements~\cite{PhysRevA.73.022328,wiese2014towards,zohar2015quantum,dalmonte2016lattice,cloet2019opportunities,Banuls:2020yw}.

Quantum information science and technology (QuIST) is currently one of the fastest growing interdisciplinary fields of research. QuIST has  brought  new  tools  and  perspectives for the calculation and computation of strongly correlated quantum systems. Understanding a dynamical process as a quantum circuit and the action of a measurement as a projection in a Hilbert space are just two instances of this quantum framework. In recent years, the scientific community has been considering several quantum technologies such as cold atoms \cite{Gross995}, trapped ions \cite{Debnath:2016tn}, or superconducting circuits \cite{blais2020circuit} as promising candidates for the realisation of a wide variety of dedicated quantum evolutions with high degree of control.

Given these advances, it is clear that the applicability of QuIST to the study of physical problems is a burning question. One possible approach is to build a multipurpose (universal, programmable) quantum computer. Yet another one  has its roots in Feynman's first intuition of quantum computers \cite{feynman1982simulating,Feynman:1986wm}: if quantum hardware able to precisely reproduce another physical quantum model exists, this would provide us with a powerful investigation tool for computing the observables of the model, and to verify or compare its prediction with the physical system. In other words, having a quantum simulator for the physical problem of interest.

Quantum simulators and quantum computers directly exploit quantum mechanical concepts such as superposition and entanglement of quantum states \cite{RevModPhys.86.153}. A fundamental reason for the exponential increase in computational power in these quantum devices is quantum entanglement, i.e., quantum correlations, among the local degrees of freedom.

A general quantum state for a set of $n$ sites, with $d$ possible quantum levels at each site requires $d^n$ complex amplitudes for its description (setting normalization aside). A classical computer will need to keep track of all these amplitudes, which means an exponential growth of memory requirements with the quantum system size. In addition to this, some quantum protocols achieve a much better scaling of computational time with the system size than any classical algorithm for the same problem \cite{Nielsen:2000ne,doi:10.1137/S0036144598347011}.

Today, the research frontier is at the edge of having universal quantum computers and quantum simulators able to perform such investigations beyond proof-of-principle analysis.  
Indeed, the quantum platforms mentioned above (cold atoms, trapped ions, superconducting circuits) are genuine quantum systems where the available experimental techniques offer an impressive degree of control together with high-fidelity measurements, thus combining two fundamental requirements for a quantum simulator. 
In this way, detailed studies and proposals have been put forward to perform quantum simulations of lattice QCD in the near and mid-term (e.g. \cite{Jordan1130,Martinez:2016qx,Mil1128,PhysRevResearch.2.013272}).
Also, light-front Hamiltonian methods to perform quantum computations of QCD matrix elements have recently been proposed as a promising alternative to equal-time lattice approaches, since they address the computation of matrix elements directly in Minkowski space-time rather than in Euclidean space-time~\cite{kreshchuk2020quantum,Kreshchuk:2020aiq}.

Let us therefore consider the conceptual requirements for quantum simulation of the quantities of interest in proton structure, such as $f_{f/P}(\xi)$ in Eq.~\eqref{eq:PDF}. We would need to encode in the quantum degrees of freedom at our disposal both matter and gauge fields. We would need to carry out measurements associated with the state $|PS\rangle$. And we need time evolution, since the Wilson line $\mathcal{U}(y)$ is non-local in time. Furthermore, we need to ensure that we are actually simulating gauge-invariant quantities. 

Recently, the simulation of dynamical gauge field has been the subject of many theoretical proposals~\cite{cloet2019opportunities,Banuls:2020yw} and the experimental realization of a scalable minimal coupling between gauge and matter field has been achieved in cold atom setups~\cite{Mil1128}. The implementation of spatial Wilson loops was considered initially in the context of topological quantum computation~\cite{Jiang:2008dw} and more recently in the context of quantum simulation of lattice gauge models~\cite{Zohar_2013,PhysRevLett.110.055302,PhysRevLett.117.240504,PhysRevD.101.034518}.

The central open problem is the one we address here: to have a quantum simulation algorithm for time-dependent quantities that are gauge invariant. Thus, as a first step towards that goal, we will consider a pure gauge model and the relevant gauge-invariant quantity: a space-time Wilson loop. Notice that this has been indeed the main stumbling block in the construction of space and time gauge-invariant quantities, and its implementation in any of the current platforms would open the floodgates of conceptual and practical developments in the topic. 

This paper is organized as follows. In Section~\ref{sec:WLdef}, we tackle the discretized construction of space-time Wilson loops along two approaches (cf.~Fig.~\ref{fig: wilson loop}), equivalent for the Abelian case: first the plaquette-based approach, valid for Abelian models, and then a link-based approach, valid for any gauge group. In the plaquette approach we present a new crucial component, the time-oriented fundamental plaquette. For the link-based approach we construct in detail the opening, propagation, and closing of the relevant lines in terms of fermionic hopping terms with gauge mediation, which preserve gauge invariance throughout. In Section~\ref{sec:WLQSim}, we discuss the quantum simulation of space-time Wilson loops in both approaches for the concrete case of a pure $Z(2)$ gauge model, in terms of circuits of quantum gates. In Section~\ref{sec:numerics}, we present a proof-of-principle computation that makes use of the algorithm. Finally, in Section~\ref{sec:conclusions}, we  discuss the main findings with a view towards further developments.

\section{Space-time Wilson loops: definition}
\label{sec:WLdef}

The primary goal of this paper is the investigation of quantum algorithms for the simulation of operators non-local in time and space, a vital step for the calculation of hadronic matrix elements. In particular, the focus will be on space-time Wilson loops. For simplicity, a pure gauge theory will be considered, in which we can develop the key features of the algorithm without introducing further complications (such as the hadronic state, matter fields, etc.). The discussion of the algorithm for other hadronic quantities, such as PDFs, will be pursued in the future.

In fact, space-time Wilson loops are very relevant matrix elements by themselves as well, in the context of nucleon structure and in particular for TMDs. For instance, a Wilson loop along both light-cone directions, the so-called ``TMD soft function'', determines the non-perturbative anomalous dimension which controls the rapidity evolution of the TMDs (see, e.g., Refs.~\cite{Echevarria:2015byo,Vladimirov:2020umg}). Also, a Wilson loop along one light-cone direction can be related to gluon TMDs at high energy, or the so-called small-$\xi$ limit (see, e.g., Refs.~\cite{Boer:2015pni,Boer:2016xqr}).

\subsubsection*{Plaquette-based space-time Wilson loop}

\begin{figure}[t]
\begin{center}
\includegraphics[width=0.41\textwidth]{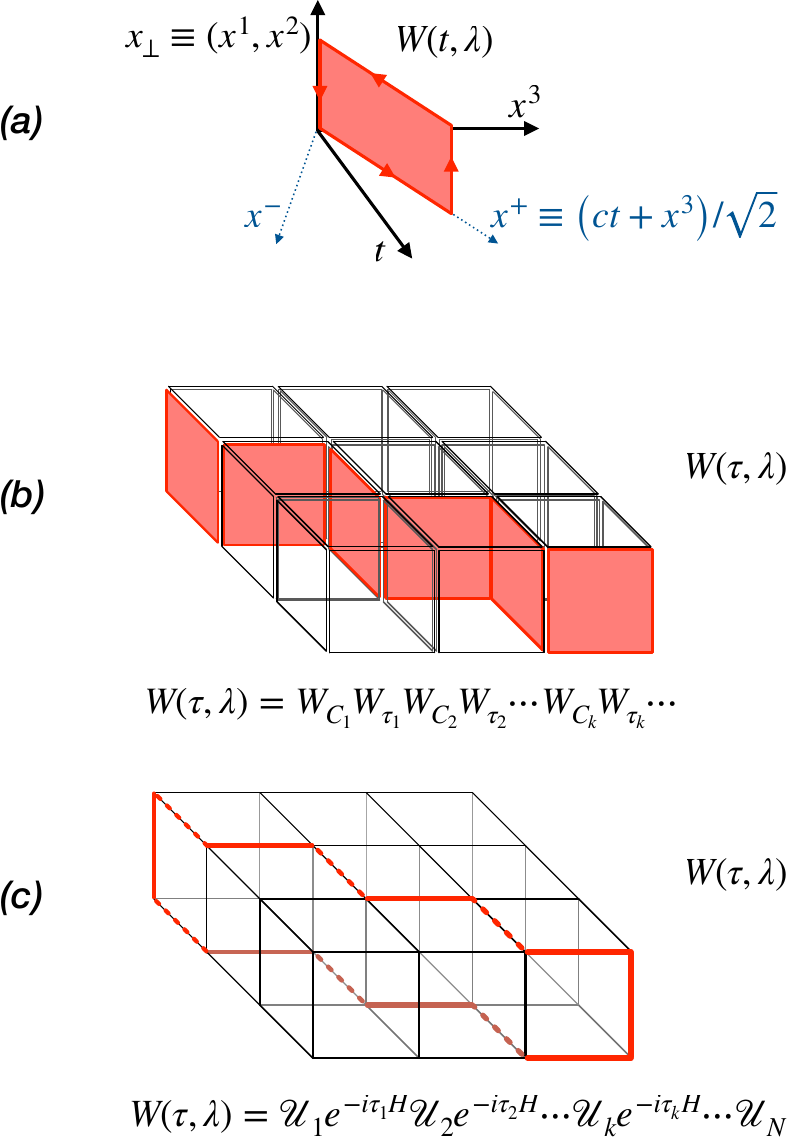}
\caption{(a) Wilson loop in the light front split in spatial and temporal planes. (b) Every space-time Wilson loop in an Abelian model can be built as the product of minimal Wilson loops in a stroboscopic evolution of spatial and temporal Wilson loops (filled squares). (c) From a link-based construction for any (Abelian and non-Abelian) Wilson loop, the stroboscopic sequence is given by spatial Wilson lines (red solid lines) and temporal Hamiltonian evolutions (red dashed lines).}
\label{fig: wilson loop}
\end{center}
\end{figure}

To quantum simulate space-time Wilson loops, first we should define these operators in an explicit gauge-invariant form in a Hamiltonian formulation. From the classical statistics definition of a gauge-invariant model~\cite{Wegner:1971dq,Wilson:1973jt,Kogut:1975zl,Kogut:1979kx}, a Wilson loop $\mathcal{W}_{\mathcal{C}} = \text{Tr} \left[ U \left( \mathcal{C} \right) \right]$ is a path-ordered unitary operator built as the product of unitary elements of the representation of the gauge group $U(e_{i})$ at each link $e_{i}$ of a closed path $\mathcal{C}$. We shall also use the name links for these unitary operators. The trace is taken in the color indices and the Wilson loop is an operator acting on the quantum Hilbert space of the corresponding degrees of freedom. As we are dealing with loops, that is, closed paths, gauge invariance is guaranteed. In order to simulate them, notice that every Wilson loop in an Abelian gauge theory can be built out of the composition of minimal Wilson loops defined in a minimal plaquette which will be our starting point.

So as to consider space-\emph{time} Wilson loops, we need to take into account the special character of the temporal direction. In the search of the definition of the quantum Hamiltonian \cite{PhysRevD.15.1128,Creutz:1985rm}, this temporal direction is taken as continuous, and the transfer matrix method provides us with the Hamiltonian. It is convenient to choose the temporal gauge, in which the links in the temporal directions are set to the identity. 

Back to the composition of a space-time Wilson loop in terms of minimal plaquettes, we see that we need two types of minimal plaquettes. First,  the pure spatial ones $\tr \left[ U(e_{1}, t) U(e_{2}, t) U(e_{3}, t) U(e_{4}, t) \right]$, where the time instant $t$ is fixed, and the four links $e_i$ form the boundary of a minimal square plaquette. Second, the temporal ones given by $\tr \left[ U(e_{i}, \tau/2 )  U(e_{i}, -\tau/2) \right] = \text{Tr} \left[ U(e_i) e^{-i\tau H} U(e_i) \right]$, where the spatial index $e_i$ is fixed and as stated above the temporal gauge has been chosen, which explains why only two unitaries appear for the four links of the plaquette, namely $e_i$ at instant $-\tau/2$, the same link at later time $\tau/2$ traversed in the opposite direction, and the two links in the temporal direction connecting the ends of the two spatial ones. For Abelian gauge models these two types of plaquettes complete the required set, and any space-time Wilson loop will be approximated by sequences of fundamental plaquettes.

Let us now make use of the temporal gauge to give explicit expressions for the temporal plaquettes in some Abelian examples:

\paragraph{Discrete Abelian $Z(2)$ gauge model.} In a $Z(2)$ gauge theory, the group element $U(e_i) = \sigma_{3}(e_i )$ is given by the third Pauli matrix. Notice that the local Hilbert space is $\mathbb{C}^2$ and, as an Abelian theory, there is no color index. The spatial plaquettes are given by $\sigma_3(e_{1},t ) \sigma_3(e_{2} ,t) \sigma_3(e_{3},t ) \sigma_3(e_{4} ,t )$ acting on $\left(\mathbb{C}^2 \right)^{\otimes 4}$. The temporal plaquettes are given by
\begin{equation}
\begin{split}
\label{z2temp}
\sigma_3(e_i, \tau/2)  \sigma_3(e_i, -\tau/2)&= \sigma_3(e_i) e^{-i\tau H} \sigma_3(e_i)\\
&= e^{-i\tau \left[ H+ 2 \sigma_1(e_i) \right]},
\end{split}
\end{equation}
with Hamiltonian
\begin{equation}
\label{eq:z2hamilt}
H=- \sum_{i} \sigma_{1}(e_i) - \lambda \sum_{\square} \sigma_3(e_1)  \sigma_3(e_2)  \sigma_3(e_3)  \sigma_3(e_4),
\end{equation}
 where $\sigma_{1}(e_i )$ is the first Pauli matrix, such that $\sigma_1 \sigma_3 = - \sigma_3 \sigma_1$, $\lambda$ is the coupling constant, and the first sum in the Hamiltonian is over all links $i$ in the lattice and the second sum is over all minimal square plaquettes $\square$. We shall further examine this Hamiltonian in section \ref{sec:numerics}.

\paragraph{Continuous Abelian \(U(1)\) gauge model.} In a $U(1)$ gauge-invariant model, the Hamiltonian is given by
\begin{equation*}
H= \sum_{i}\frac{g^{2}}{2} L^{2}(e_i) - \frac{1}{g^{2}} \sum_{\square} \text{Re} \left[ U(e_1) U(e_2) U(e_3) U(e_4) \right],
\end{equation*}
where $g$ is the coupling constant and $[L(e_i) , U(e_i)] = U(e_i)$ are conjugate variables. $U(e_i)$ is the group element and $L(e_i)$ the electric field. In this case, the minimal temporal Wilson loop reads
\begin{equation}
\label{u1temp}
\text{Tr} \left[ U^{\dagger}(e_i) e^{-i\tau H} U(e_i) \right]  = e^{-i\tau \left[H + \frac{g^{2}}{2}  \left( 2L(e_i)+1\right) \right]}.
\end{equation}

Thus, as we have seen in the examples of Eqs.~\eqref{z2temp} and \eqref{u1temp}, elementary temporal Wilson loops appear as unitary temporal evolution, with the Hamiltonian $H$, obtained from the transfer matrix method, being modified by the additions of an operator $O\left(e_m \right)$ localized on the relevant link $e_m$, i.e., $\mathcal{W}_{\tau_m} =e^{-i\tau_m \left[ H + O\left(e_m \right) \right]}$. This structure is universal, while the concrete additional operator $O(e_m)$ is model dependent

Restating our objective of constructing Wilson loops, spatial Wilson loops $\mathcal{W}_{\mathcal{C}_n}$ correspond to path-ordered unitary operators $U\left(e_i\right)$ on contiguous links forming a closed loop $\mathcal{C}_n$, i.e., $\mathcal{W}_{\mathcal{C}_n} = \mathcal{P}_{\otimes_{e_{i} \in \mathcal{C}_n}} U\left(e_i\right)$ \cite{Jiang:2008dw,PhysRevLett.117.240504,PhysRevD.101.034518}. Thus the complete space-time Wilson loop can be stroboscopically decomposed in spatial and temporal Wilson loops given by
\begin{equation}
\label{strob}
\mathcal{W} = \mathcal{W}_{\mathcal{C}_1} \mathcal{W}_{\tau_1} \mathcal{W}_{\mathcal{C}_2} \mathcal{W}_{\tau_2} \cdots \mathcal{W}_{\mathcal{C}_k} \mathcal{W}_{\tau_k} \cdots,
\end{equation}
and in the next section we shall examine its realization in terms of quantum gates.

\subsubsection*{Link-based space-time Wilson loop}

\begin{figure}[ht!]
\begin{center}
\includegraphics[width=0.5\textwidth]{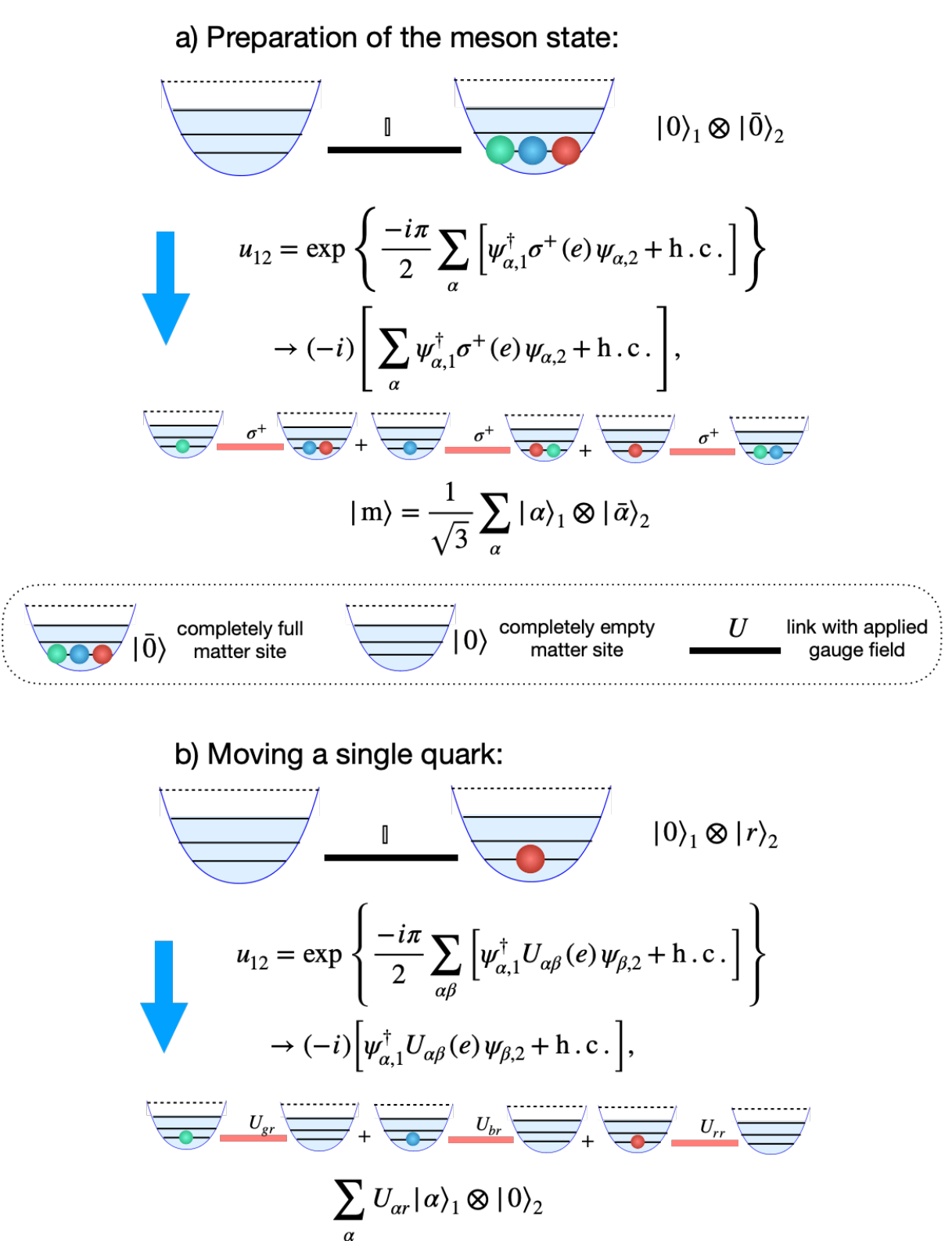}
\caption{Preparation of a meson state and minimal quark transport in the lattice. (a) Starting from a completely empty and completely full state $|0 \rangle \otimes |\bar{0} \rangle$, a meson state $|\text{m} \rangle  \equiv \frac{1}{N^{1/2}}\sum_{\alpha=1}^N |\alpha, \bar{\alpha} \rangle$ is built (b) By sequentially applying the parallel transport of a single quark in the lattice, the link-based Wilson loop is built.}
\label{fig:linkWilson}
\end{center}
\end{figure}

\begin{figure*}[ht!]
\begin{center}
\includegraphics[width=0.7\textwidth]{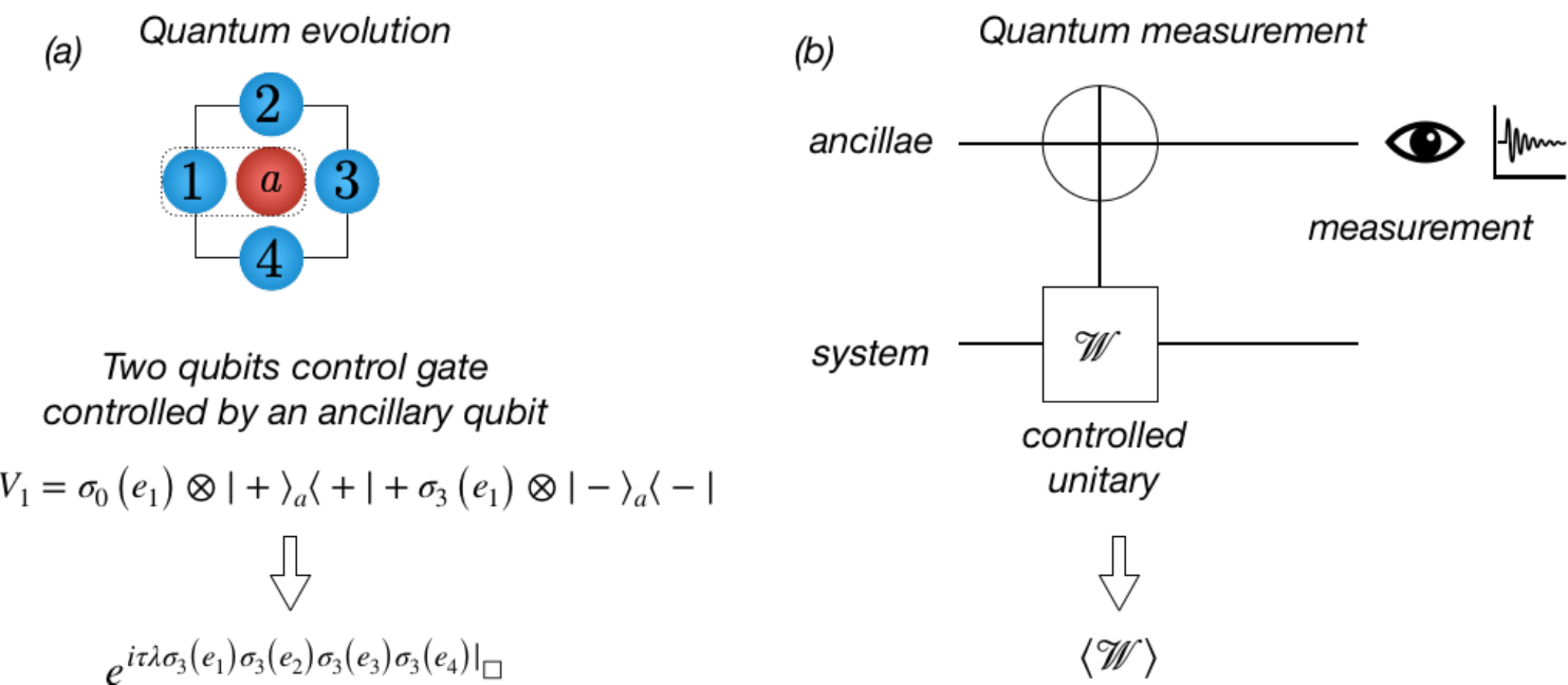}
\caption{Quantum controlled gates as the core for the quantum evolution and the quantum measurement. (a) A two qubit quantum controlled gate is the basis of plaquette or magnetic interactions in the gauge-invariant model (b) Entangling a quantum many-body system with an ancillary qubit using a controlled Wilson gate is the basis for measuring any Wilson loop in the many-body system}
\label{fig:quantum simulation}
\end{center}
\end{figure*}

In a second approach (see Fig.~\ref{fig:linkWilson}), we explicitly build a non-Abelian space-time Wilson-loop operator by 1) inserting a quark-antiquark pair in adjoining sites that are part of the path of the Wilson loop; 2) parallel transport of the  quark  and antiquark quantum states in opposite directions along the loop; and 3) annihilation of the quark-antiquark pair to close the loop \cite{Zohar_2013,PhysRevLett.110.055302}. Each of these three set of actions is achieved, in the simulation, by acting on reference states of gauge and matter with gauge-invariant hopping operators. As they are hopping terms, they pertain to the spatial part of the Wilson loop. Thus, to have a space-time Wilson loop we need to incorporate in this description the temporal links. These are simply achieved by temporal evolution with the transfer matrix Hamiltonian of the gauge model, with no evolution for the quarks and antiquarks. That is to say, the matter fields are non-dynamical and ancillary.

In other words, by quark here we mean that an ancillary degree of freedom has been loaded with a fermion,  $|\alpha\rangle = \psi^{\dagger}_{\alpha} |0\rangle$, where we use the label $\alpha$ as in the colour indices of the unitaries of the group. The initial action of the construction is loading a couple of adjoining sites along the path with an $N$-quark singlet state of $SU(N)$, thus totally antisymmetric, that can be understood as a maximally entangled state of a quark-antiquark pair (\emph{meson}), 
\begin{equation}
\label{eq:mesonstate}
|\text{m} \rangle \equiv \frac{1}{N^{1/2}}\sum_{\alpha=1}^N |\alpha (1), \bar{\alpha} (2)\rangle,
\end{equation}
where
\begin{equation*}
|\bar{\alpha} \rangle= \frac{1}{(N-1)!} \sum_{\beta_i} \epsilon^{\alpha \beta_1 \cdots \beta_{N-1}}  \psi^{\dagger}_{\beta_1} \cdots \psi^{\dagger}_{\beta_{N-1}} |0\rangle,
\end{equation*}
with $\epsilon^{\alpha \beta_1 \cdots \beta_{N-1}}$ being the totally antisymmetric tensor 
and $N$ the number of colors, is understood as an "antiquark" with color $\alpha$. 

The labels 1 and 2 in the meson state \eqref{eq:mesonstate} correspond to the sites where the quark and antiquark are located. In this context, sites are the endpoints of links, and will conceptually be locations for the ancillary matter fields. In the actual process of simulation the ancillary fermionic states could be coded in a different physical location and in fact be reused to describe different sites of the simulation. Leaving that for the implementation, in what follows we will be working with fermionic operators acting on different sites, the ancillary matter sites or lattice vertices, where the matter field operators $\psi_{\alpha,j}$, with color index $\alpha$ and spatial index $j$, live. Their statistics is fermionic, i.e., $\{ \psi_{\alpha,j} ,\psi^\dagger_{\beta,k} \}=\delta_{\alpha, \beta}\delta_{j,k}$, and there is a local reference state (\emph{vacuum} or \emph{empty singlet}) $|0(j) \rangle$ such that $\psi_{\alpha,j} |0(j) \rangle =0$, $\forall \alpha$, $j$; there is a second reference state (\emph{full singlet}) $|\bar{0}(j) \rangle= \frac{1}{N!} \sum_{\beta_i} \epsilon^{ \beta_1 \cdots \beta_{N}}  \psi^{\dagger}_{\beta_1} \cdots \psi^{\dagger}_{\beta_{N}} |0(j)\rangle$ such that $\psi^\dagger_{\beta,j} |\bar{0}(j) \rangle =0$, $\forall \beta$, $j$; notice that $ |\bar{\alpha} \rangle = \psi_{\alpha} |\bar{0} \rangle$. 
Note that the antisymmetry of the creation and annihilation operators is only needed locally. The process of creating a link-based Wilson loop is based on single-particle physics where the statistics of the operators is not relevant.

Other than the two sites referenced in the meson state \eqref{eq:mesonstate}, all ancillary matter sites are initialised to a reference state, either $|0(j)\rangle$ or $|\bar{0}(j)\rangle$. In fact, the meson state $|\mathrm{m}\rangle$ will be obtained by applying on the total vacuum state $|0(1)\rangle|\bar{0}(2)\rangle$ the creation/annihilation hopping term of Eq. \eqref{eq:openclosehopping}, that we will describe shortly. For the time being, consider the initial meson state as given, and we shall now propagate the quark and antiquark states along spatial links, until a point where a temporal evolution is necessary. When the discretization of the Wilson loop that is desired requires again spatial links, the propagation of the quark and antiquark resumes as we now describe.

The (spatial) propagation of the quark state proceeds by acting on a succession of $|0(j)\rangle$ states sequentially with gauge-invariant hopping terms of the form 
\begin{equation}
\begin{split}
u_{12} = &\exp{\left\{\frac{-i \pi}{2}   \sum_{\alpha \beta}\left[ \psi^\dagger_{\alpha,1} U_{\alpha \beta}\left( e \right)  \psi_{\beta,2} + \text{h.c.} \right]  \right\}} \\
\to& (-i)  \left[ \psi^\dagger_{\alpha,1} U_{\alpha \beta}\left( e \right) \psi_{\beta,2} + \text{h.c.} \right],
\end{split}
\label{hopterm}
\end{equation}
where $e$ corresponds to the link between sites $1$ and $2$ along the minimal Wilson line, and the last assignment is valid in the single quark sector. We have to identify on which part of the Wilson line we shall propagate the quark and on which one the antiquark. In the quark propagation part all the matter sites will be initially set to the $|0\rangle$ state, while those deemed to support the antiquark will be prepared in the $|\bar{0}\rangle$ state.

For definiteness,  let  us assume an initial spatial part of the Wilson line to be of  odd length $L+1$. We initialize two ancillary sites in the entangled singlet state, $\sum_{\alpha} |\alpha (A_{L/2}), \bar{\alpha} (B_{L/2})\rangle$. We use the notation $A_{L/2}$ and $B_{L/2}$ to signal that we will be moving out of the center at $L/2$ and take this entanglement to the two boundaries of the line. The links enumerated with 1 to $L/2-1$ will be carrying the quark, and thus set to $|0\rangle$, while those branching out from $B_{L/2}$ to $L$, corresponding to the spatial propagation of the quark, will be initialised to $|\bar{0}\rangle$.

The quark will  move out of the center because of the action of  hopping term \eqref{hopterm} on this initial state with the central meson, as follows:
\begin{equation*}
\begin{split}
&u_{L/2-1,L/2} |0 (L/2-1)\rangle  \otimes |\alpha (A_{L/2}), \bar{\alpha} (B_{L/2})\rangle \\
&=|\beta (A_{L/2-1})\rangle U_{\beta \alpha}(e_{L/2-1})  \otimes |0 (L/2) , \bar{\alpha}(B_{L/2}) \rangle.
\end{split}
\end{equation*}

Iterating this process with contiguous links towards the initial point of the line, a Wilson line operator of the form $|\alpha (A_{1}) \rangle U_{\alpha \beta}(e_{1})U_{\beta \gamma}(e_{2}) \cdots U_{\mu \nu}(e_{L/2-1}) |\bar{\nu}(B_L/2) \rangle$ is built, where all the internal color indices are contracted in a path order product of parallel transporters, the initial and final color indexes are contracted with the ancillary matter sites and the intermediate matter sites are uncoupled in a product state of empty states.

In a similar way, the \emph{anti-quark} state $|\bar{\nu}\rangle$ can be parallel transported towards the end of the line, if the contiguous ancillary \emph{anti-matter} sites are initialized to the full reference states $|\bar{0}(j)\rangle$, $\forall j > {L/2}$. Then, the complete Wilson line operator will be
\begin{equation}
\begin{split}
\mathcal{U}&(A_{1},B_L) = \frac{1}{N^{1/2}} \sum_{\alpha\beta \cdots \mu \nu \omega \cdot \theta \phi}|\alpha (A_{1}) \rangle |\bar{\phi}(B_L) \rangle \\
& U_{\alpha \beta}(e_{1}) \cdots U_{\mu \nu}(e_{L/2-1}) U_{\omega \nu}^*(e_{L/2}) \cdots U_{\phi \theta}^*(e_{L-1}) \\
=& \frac{1}{N^{1/2}} \sum_{\alpha \phi}|\alpha (A_{1}) \rangle \mathcal{U}_{\alpha \phi}(e_{1}, \cdots ,e_{L-1} )  |\bar{\phi}(B_L) \rangle
\end{split}
\end{equation}

This operator acts in a space slice of constant time \(t\) in the space-time. Then we will evolve the system with the unitary operator $e^{-i\tau H}$ for a time interval $\tau$. The Hamiltonian in our construction only involves the gauge degrees of freedom and the ancillary matter degrees of freedom have no dynamics, as stated previously. Their only role is as registers of the color indices. 

After the final spatial Wilson line, that leaves the first and last matter sites as neighbours, and in order to complete the loop, we need a different hopping term between the first and the last matter sites. This comes about because in the Wilson loop there is a clear path ordering with a definite orientation. We introduce an ancillary qubit such that we have at our disposal the following hopping term
\begin{equation}
\label{eq:openclosehopping}
u_{B_L1} = \exp{\left\{ \frac{-i \pi }{2 N^{1/2}} \sum_{\beta}\left[ \psi^\dagger_{\beta,B_L}  \sigma^+  \psi_{\beta,A_1} + \text{h.c.} \right]  \right\}}.
\end{equation}

This will be applied to a Wilson lines with ancillary degrees of freedom $|\bar{\phi}(B_L) \rangle |\downarrow \rangle |\alpha (A_1) \rangle$, where the state $|\downarrow\rangle$ is the extra qubit degree of freedom such that $\sigma^+ |\downarrow \rangle = |\uparrow\rangle$. In fact, we can understand this additional qubit as a local $U(1)$ gauge element, thus justifying our understanding of this term as hopping with a $U(1)$ mediator.

Specifically for the case at hand, the action of this operator on the matter sites couples just two quantum states:
\begin{equation*}
\begin{split}
u_{B_L1} \frac{1}{N^{1/2}} & \sum_{\gamma} |\bar{\gamma}(B_L) \downarrow \gamma (A_1) \rangle\\
&= i (-1)^{N}  |\bar{0}(B_L) \uparrow 0 (A_1) \rangle \, .
\end{split}
\end{equation*}
If we measure the ancillary qubit, the probability to be in the \(|\uparrow \rangle\) state is proportional to the Wilson loop, $P_{\uparrow}=\Big| \frac{\text{Tr} ( \mathcal{U} ) }{N}  \Big|^2$, with $\text{Tr}$ the standard trace over color indices. Thus, when the outcome of our measurement is $\uparrow$, we are assured that our Wilson loop has been constructed and is applied to the gauge links in the many-body quantum state, non-destructively \cite{PhysRevD.65.065022}.

Notice furthermore that if the central pair of sites in the first spatial part of the line were adjoined with the same ancillary qubit, and the latter were prepared in the $|\uparrow\rangle$ state, then the action of \eqref{eq:openclosehopping} would create the desired meson state. As we see, the link approach we have presented requires, other than the Hamiltonian for the gauge degrees of freedom, of two types of hopping terms. The second kind of hopping, Eq. \eqref{eq:openclosehopping}, requires an additional qubit, that can be reused with no additional overhead. The first hopping term, Eq. \eqref{hopterm}, acts just on the quark/antiquark vacuum and the previously evolved state.

\section{Space-time Wilson loops: Quantum simulation}
\label{sec:WLQSim}

\subsubsection*{Plaquette-based space-time Wilson loop}

Every step in the stroboscopic decomposition of Eq. \eqref{strob}, be it $\mathcal{W}_{\mathcal{C}_n}$ or $\mathcal{W}_{\tau_m}$, is gauge invariant by construction. This digital (stroboscopic) approach renders any possible Abelian gauge symmetry amenable to quantum simulation. The decomposition presented in Eq. \eqref{strob} leads to products of local or electric terms and minimal plaquette or magnetic terms. We therefore need implementable simulations of these minimal gauge-invariant operators.

For clarity, in what follows, we will describe fully this algorithm for the simplest pure $Z(2)$ gauge-invariant model \cite{Weimer:2010tv,PhysRevLett.118.070501,Schweizer:2019km}. 

A reasonable minimal demand for the physical implementation of this process is the availability of two types of unitary gates: i) local ones of the form $e^{i \tau  \sigma_{1} (e_i )}$ for some given time interval $\tau$, and ii) collective ones that are the exponentiation of plaquette operators $\sigma_3(e_1 ) \sigma_3(e_2 ) \sigma_3(e_3 ) \sigma_3(e_4 )|_{\square}$ for a time interval $\tau$. 

Given this set, the relevant temporal plaquette Eq.~\eqref{z2temp} is at our disposal as well by means of a Trotter approximation, while the spatial plaquettes correspond to the particular value $\tau = \pi /2$ of the collective unitaries.

In actual fact, it is rather unlikely that we will have a four-link unitary operator at our disposal, so it behoves us to provide a constructive method for it (for illustration see Fig.~\ref{fig:quantum simulation}). In particular, it can be achieved by the action of a two-qubit gate, controlled by an ancillary qubit (denoted by subindex $a$), and acting on the $e_i$ link,  $V_{i}=\sigma_{0} (e_i ) \otimes|+ \rangle_{a} \langle + | + \sigma_{3}(e_i ) \otimes|- \rangle_{a} \langle - |$, where $|\pm \rangle$ are eigenstates of $\sigma_1 |\pm \rangle = \pm |\pm \rangle$, and $\sigma_{0}$ is the identity. For a given spatial loop, one applies an ordered sequence of these two qubit gates with a common ancilla,  from the first to the last qubit around the closed loop, $V_{123 \cdots n,a} = V_{1} V_{2} V_{3} \cdots V_{n} = |+ \rangle_{a} \langle + | + \sigma_{3}\left(1\right) \sigma_{3}\left(2\right) \sigma_{3}\left(3\right) \cdots \sigma_{3}\left(n\right) |- \rangle_{a} \langle -|$. 

In this manner, were the ancillary qubit prepared in the state $|-\rangle_{a}$, then
\begin{equation}
\label{plaq}
V_{123 \cdots n,a} |-\rangle_{a}= \mathcal{W}_{\mathcal{C}} |-\rangle_{a} = \mathcal{P}_{\otimes_{e_{i} \in \mathcal{C}_n}} \sigma_3 \left( e_i \right) |-\rangle_{a},
\end{equation}
thus constructing the spatial Wilson loop operator.

We also require the exponentiated form of a minimal loop for a given time $\tau$. We start by preparing the ancillary qubit in the state $|\downarrow \rangle_{a}$. Here, $|\uparrow / \downarrow \rangle_{a}$ are defined by $\sigma_3 |\uparrow / \downarrow \rangle = + / -|\uparrow / \downarrow \rangle$. Next we apply the unitary operator $V_{1234,a}$, followed by the local evolution $e^{-i \lambda \tau \sigma_{3}\left( a \right) }$ and finally $V^{\dagger}_{1234}$, i.e., 
\begin{equation}
\label{plaqdyn}
\begin{split}
V^{\dagger}_{1234,a} e^{-i \lambda \tau \sigma_{3}\left( a \right) } &V_{1234,a} |\uparrow \rangle_{a} = \\
& e^{i \tau \lambda \sigma_3(e_1 ) \sigma_3(e_2 ) \sigma_3(e_3 ) \sigma_3(e_4 )|_{\square} } |\downarrow \rangle_{a}.
\end{split}
\end{equation}
In this manner, we have achieved both the spatial \eqref{plaq} and temporal \eqref{plaqdyn} plaquettes, as promised.

\subsubsection*{Link-based space-time Wilson loop}

The link-based approach presented in the previous section is applicable to both Abelian and non-Abelian gauge invariant models. Nonetheless, for the sake of definiteness, we shall again present the quantum simulation for the pure $Z(2)$ gauge invariant model, now in this approach.

Our starting point is the assumption that we have access to two types of interactions: i) as in the previous section, local ones of the form $e^{i \tau  \sigma_{1} (e_i )}$ for some given time interval $\tau$, and ii) collective ones that involve the matter-gauge interaction of the form $H=\sigma^+_j \sigma_3(e_j) \sigma^-_{j+1} + \text{h.c.}$, where the matter field plays an ancillary role in the whole process and presents no dynamics on its own.   

With the matter-gauge interaction acting for a time $\pi/2$ on the matter states $|\downarrow_j \uparrow_{j+1}\rangle$, the result is the minimal Wilson line $|\uparrow_j\rangle \sigma_3(e_j) |\downarrow_{j+1}\rangle$. Iterating the process, any Wilson line on a time slide $t$ can be built. For instance, the minimal plaquette Wilson loop is built by the action of the matter-gauge interaction in a closed loop around a plaquette $\sigma_3(e_1) \sigma_3(e_2) \sigma_3(e_3) \sigma_3(e_4)|_{\square}$. Notice that the actual state of the ancillary matter degrees of freedom are completely decoupled from the Wilson loop. Finally, the dynamics of a single plaquette follows the description of Eq.~\eqref{plaqdyn}, thus completing the necessary set of links and dynamics.
 
\subsection*{Non-demolition measurement of a space-time Wilson loop}

Once the object of interest has been built, in our case the space-time Wilson loop, it is now necessary to obtain information from and about it. Let us put forward two schemes pertaining the quantum simulation of space-time Wilson loops.

(i) Local measurement in the local basis that diagonalizes the ``electric" field, i.e., in a gauge-invariant basis. In this way, the experiment has to be repeated several times to obtain the distribution of the electric field in the lattice. This distribution is determined by the Wilson loop operator and the initial states of the gauge degrees of freedom.

(ii) Alternatively, a quantum non-demolition measurement of a spatial-temporal Wilson loop is possible, using a controlled Wilson loop with an ancillary qubit (see Fig.~\ref{fig:quantum simulation}). 

Let us define, for a general unitary operator $U$, a controlled version of it as $c\text{-}U=\mathbb{I}_{\text{syst}} |\downarrow\rangle_{a} \langle \downarrow| + U_{\text{syst}} |\uparrow \rangle_{a} \langle \uparrow |$. Its action on an arbitrary state of the system $|\psi\rangle_{\text{syst}}$ and the state $|+\rangle_{a} = \frac{1}{\sqrt{2}} \left( | \uparrow \rangle_{a} + | \downarrow \rangle_{a} \right)$ results in $c\text{-}U |\psi\rangle_{\text{syst}} |+\rangle_{a} = \frac{1+ U_{\text{syst}} }{2}  |\psi\rangle_{\text{syst}} |+\rangle_{a} \frac{1- U_{\text{syst}} }{2}  |\psi\rangle_{\text{syst}} |-\rangle_{a} $, whence, on measuring the probability of obtaining the state $+$, we obtain the expectation value of the Wilson loop
\begin{equation}
\label{prob}
p_{+}=  _{\text{syst}}\langle \psi | \frac{2+U_{\text{syst}} + U^{\dagger}_{\text{syst}} }{4} |\psi\rangle_{\text{syst}} 
\end{equation}

In view of this, we now face the problem of building the controlled spatial-temporal Wilson loop. Let us first consider local terms $e^{i\tau  \sum_{ \vec{r}} \sigma_{1}(\vec{r})}$, for which the system-ancilla interaction given by $H^{\Gamma}_{\text{syst-}a}=\frac{\sigma_{3}(a) +1 }{2} \sigma_{1}(\vec{r})$, when acting during an interval of time $\tau$, results in the gate $c\text{-}U^{\Gamma}=\mathbb{I}_{\text{syst}} |\downarrow\rangle_{a} \langle \downarrow| + e^{i\tau  \sum_{ \vec{r}} \sigma_{1}(\vec{r})} |\uparrow \rangle_{a} \langle \uparrow |$. The spatial Wilson loops can be achieved in a similar way, with an interaction of the form $H^{\mathcal{C}}_{\text{syst-}a}=\frac{\sigma_{3}(a) +1 }{2} \sum_{\vec{r} \in \square} \sigma_{3}(\vec{r})$ for a time $\tau=\frac{\pi}{2}$, in which case the resulting quantum gate is $c\text{-}U^{\mathcal{C}}=\mathbb{I}_{\text{syst}} |\downarrow\rangle_{a} \langle \downarrow| + \otimes_{\vec{r}\in \mathcal{C}} \sigma_3 \left( \vec{r} \right) |\uparrow \rangle_{a} \langle \uparrow |$. As to the plaquette interaction, we can achieve it sequentially with two ancillary qubits, i.e.,
\begin{equation*}
\begin{split}
c\text{-}U^{\square}&=\mathbb{I}_{\text{syst}} |\downarrow\rangle_{a} \langle \downarrow| + e^{i \tau \lambda \sum_{\square} \sigma_3 \sigma_3 \sigma_3 \sigma_3|_{\square} }  |\uparrow \rangle_{a} \langle \uparrow | |\downarrow \rangle_{b} \\
=&V^{\dagger}_{1234,b} \left[ \mathbb{I}_{\text{syst}} |\downarrow\rangle_{a} \langle \downarrow| + e^{-i\tau \sigma_{3}\left( b \right) }  |\uparrow \rangle_{a} \langle \uparrow | \right] V_{1234,b} |\downarrow \rangle_{b}
\end{split}
\end{equation*}

In summary, after the application of the sequence of controlled unitaries we have presented here,  measurement of the ancillary qubit provides Eq.\eqref{prob}, the expectation value of the Wilson loop of interest.

\subsection*{Scaling}
In any quantum simulation it is crucial to have at least an estimate of the number of qubits and gates that it requires. Even though this information does not suffice to determine its viability, since usually some gates will be more prone to error and will become a bottleneck for its application, it is relevant to assess its usefulness. In the case at hand, there are more aspects to consider, as we now analyze. There are two main differences between the plaquette and the link approaches in this regard. 

In the plaquette proposal we have put forward for Abelian gauge theories  we need a few ancillary qubits that can be systematically reused, so in fact when it comes to the number of qubits it is determined by the coding of the Abelian degrees of freedom plus a rather small ancillary overhead. Gatewise, the number of different types of gates that need implementing is also moderate, as noted in this section. As to the actual number of gates, it will generally scale as $L^2$, where $L$ is the total length of the Wilson loop under investigation, in an area scaling law. 

Passing now to the link-based approach, its  ancillary content is somewhat more sophisticated, and  dependent on the number of colours. On the other hand, the ancillary fermionic degrees of freedom need not be at our disposal for all vertices of the lattice. The system that codes a matter site can be reused straightforwardly with a reset to $|0\rangle$ or $|\bar{0}\rangle$, as the case might be. In the systematics of quark-antiquark propagation we actually only need four matter sites, in total. To these we should add a qubit for the opening/closing hopping unitary, as well. The coding of the non-Abelian gauge degrees of freedom will also be more demanding in terms of qubits than the Abelian case, if a digital coding is desired or available. The fact that the closing of the Wilson loop is probabilistic will add runtime to a simulation as well, certainly. To make up for these drawbacks, the link-based approach has the definite advantage that the number of gates will scale \emph{linearly} with the length of the Wilson loop, $L$. In order to estimate fully the number of gates the degree of Trotterization  \cite{Lloyd1073,childs2019theory} would need to be determined, but the overall $L^\alpha$ scalings we have indicated here will be controlling.

\section{Proof of principle simulation}
\label{sec:numerics}

\begin{figure}[ht!]
\begin{center}
\includegraphics[width=0.45\textwidth]{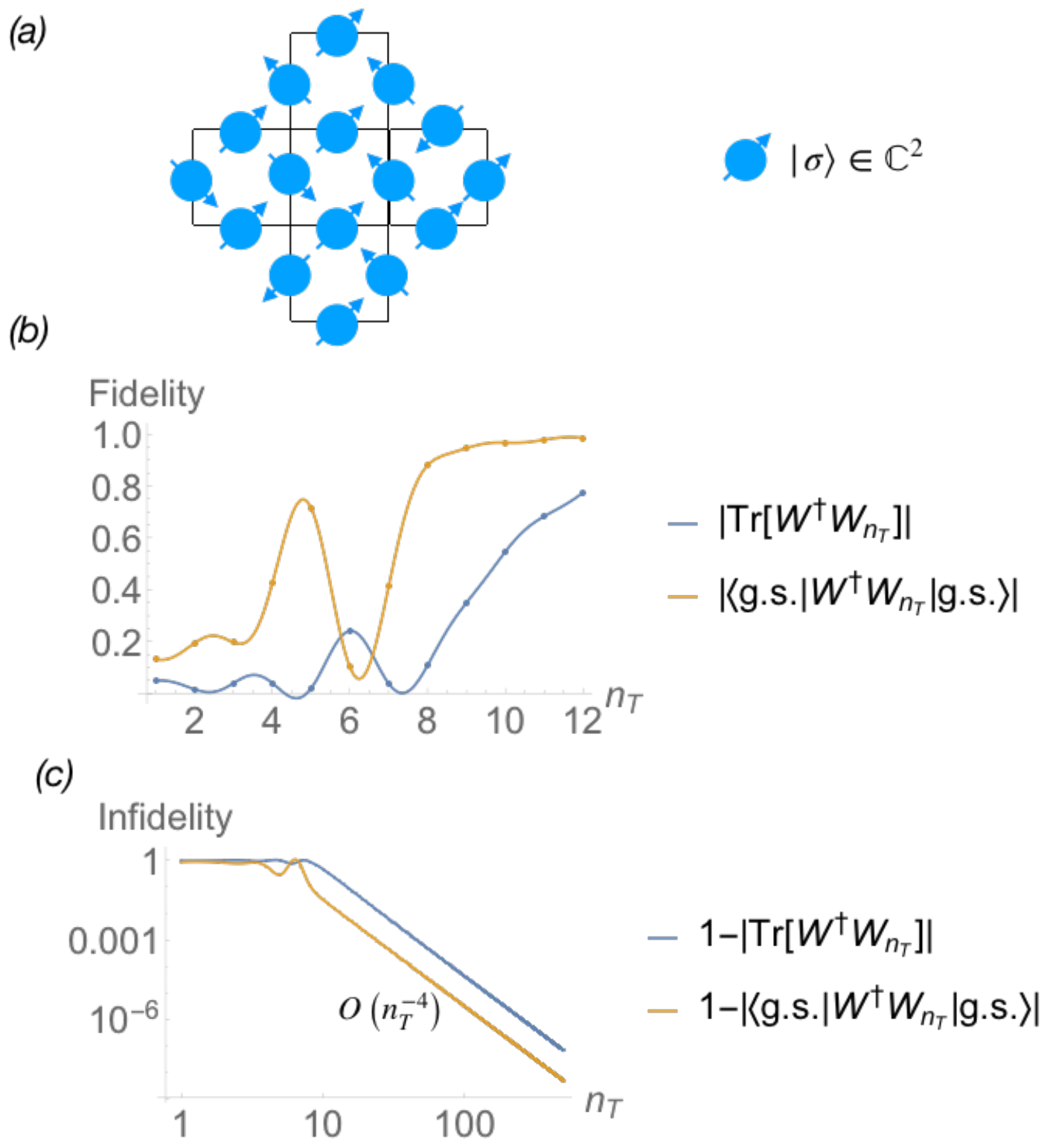}
\caption{a) Lattice setup for the quantum simulation of a minimal space-time Wilson loop. At every link of the lattice, there is a qbit. The total Hilbert space is of dimension $2^{16}$, the gauge invariant or physical Hilbert space is of dimension $2^5$. b) Minimum Trotter steps $n_T$ such that the fidelity of the Wilson loop is closed to one. c) Numerical results for the operator infidelity and ground state infidelity, where the error in the Wilson loop operator scales with the number of Trotter steps $n_T$ like a power law, with an exponent $n_T^{-4.010(1)}$ for the operator infidelity and $n_T^{-3.9997(2)}$ for the ground state infidelity.}
\label{fig:numerical results}
\end{center}
\end{figure}

In this last section, we explicitly compute the space-time Wilson loop shown in Fig.~\ref{fig: wilson loop} and we check the effect of the Trotterization \cite{Lloyd1073,childs2019theory} in the dynamics of the quantum simulation of a $Z(2)$ space-time Wilson loop.

The quantum Hamiltonian of a pure $Z(2)$ gauge-invariant model is given by Eq.~\eqref{eq:z2hamilt}, that we present here again for reference
\begin{equation*}
\begin{split}
H&= H_{\text{el}} + \lambda H_{\text{mag}} \\
&=- \sum_{i} \sigma_{1}(e_i) - \lambda \sum_{\square} \sigma_3(e_1)  \sigma_3(e_2)  \sigma_3(e_3)  \sigma_3(e_4),
\end{split}
\end{equation*}
where $\lambda$ is the coupling constant and $H_{\text{el}}$ ($H_{\text{mag}}$) corresponds to the electric (magnetic) interaction in the lattice gauge model. The local constraint around every vertex $+$ of the lattice due to the gauge symmetry reads as $\bigotimes_{e_i \in +} \sigma_1(e_i) |\text{phys}\rangle = |\text{phys}\rangle$. Alternatively, at each full vertex the projector onto physical states $P_{\mathrm{phys}}=\left({1}+\bigotimes_{e_i \in +} \sigma_1(e_i)\right)/2$ is of rank 8, and halves the number of degrees of freedom. Notice that in the full lattice one of the vertex projectors is redundant. It is well known \cite{Wegner:1971dq} that the phase diagram of this model has two phases: for $\lambda \ll 1$ the system is in a confined  phase, while for $\lambda \gg 1$ in a deconfined one. At $\lambda \sim 1$, the model is critical, the mass gap goes to zero, and the correlation length to infinity in the thermodynamical limit.

For the numerical simulation, we assume a minimal setup with 16 qubits in a ``cross" configuration as  shown in Fig.~\ref{fig:numerical results}. Due to the gauge constraints around every vertex, the gauge-invariant or physical Hilbert space is of dimension $2^5$. We set the value of the coupling constant to $\lambda = 10$, as the system is thus close to its critical point. 

We are targetting the simulation of light-front physics. As a consequence it is convenient to select the spatial and temporal lattice spacings to be of the  same order. By this we mean that one time step entails  the evolution of the quantum Hamiltonian for a continuous time  interval, $\tau$,  of the same order as the lattice spacing, $\tau \sim 1$. Thus, each one time step is obtained by the application of the unitary $e^{-i H} = e^{-i \left( H_{\text{el}} + \lambda H_{\text{mag}} \right) }$.

This fundamental one time step evolution will be Trotter expanded as  
\begin{equation}
\label{eq:Trotter}
e^{-i H} \simeq \left[e^{-i H_{\text{el}}/2{n_T}} e^{-i\lambda H_{ \text{mag}}/{n_T}} e^{-i H_{\text{el}}/2{n_T}} \right]^{n_T}\,,
\end{equation} with $n_T$ Trotter steps, in the second order symmetric Trotter--Suzuki approximation. We are assuming homogeneous Trotterization for all elementary time evolutions. In other words, $n_T$ is the same for all evolutions from one time slice to the next. In more refined implementations of the algorithm, an adaptive Trotterization might prove advantageous. Given a Wilson loop $\mathcal{W}$, we denote as $\mathcal{W}_{{n_T}}$ its Trotterized version, i.e., the one obtained from the plaquette construction where the elementary evolution is decomposed according to Eq.~\eqref{eq:Trotter}.

We will consider two figures of merit for the quality of the quantum simulation of the space-time Wilson loop: \emph{i)} the operator fidelity between the continuous time operator and the Trotterized one $\big|\text{Tr}\left[ \mathcal{W}^\dagger \mathcal{W}_{{n_T}} \right] \big|$, with normalized trace ($\text{Tr}\left[1\right]=1$); and \emph{ii)} the ground-state fidelity $\big|\langle \text{g.s.} | \mathcal{W}^\dagger \mathcal{W}_{{n_T}} | \text{g.s.} \rangle \big|$, where $|\text{g.s.}\rangle$ corresponds to the ground state of the quantum Hamiltonian with a given coupling $\lambda$. Notice that the operator fidelity and the ground-state fidelity will in general be different. For a general investigation of the Wilson loop, the operator fidelity will be more relevant; for the explicit construction at hand, in which we have a reference state (in particular, the ground state of the gauge model Hamiltonian), the second figure of merit will be of interest.

These figures of merit will be  applied to investigate two questions. First, a numerical estimation of the range of sensible number of Trotter steps required for a reliable simulation. Second, the general scaling of fidelities with the number of Trotter steps. As is well known, the estimate bounds for fidelities in the Trotter approximation \cite{childs2019theory} are notoriously not sharp in general (see for instance how interference can suppress error so that it falls below the standard estimate in \cite{Childs2020}). Thus it is of interest to investigate whether or not the asymptotic behavior for infidelities (one minus the fidelity) deviates from the standard estimate, namely $O\left(n_T^{-4}\right)$.

We therefore carry out numerical estimations of these  figures of merit in the cross configuration we have described, as depicted in Fig.~\ref{fig:numerical results}. First, we depict the fidelities as a function of $n_T$, the number of Trotter steps for each evolution in the discrete time interval $\tau$. As usual, the fidelities are not a simple linear function in $n_T$, with a peculiar dip in the ground state fidelity for intermediate values. We see that  for $n_T \sim 9$, the ground state fidelity is $\big|\langle \text{g.s.} | \mathcal{W}^\dagger \mathcal{W}_{n_T} | \text{g.s.} \rangle \big| \sim 0.95$, which can be competitive for many applications. One should bear in mind in this respect  that this is the \emph{total} fidelity for a highly nonlocal operator. Second, we depict the asymptotic behavior of the infidelity with the number of Trotter steps. In this second scenario, both the ground state and operator infidelities behaviors are similar, decreasing the errors in the Wilson-loop operator with a power-law dependence $O(n_T^{-4})$, as expected for a second order Trotter--Suzuki approximation. Notice that the scaling regime appears already in the first decade, with a transient that will be, in general, dependent on implementation.

Finally, the link based approach will not be of particular advantage in this proof-of-principle example, with 14 links and 6 plaquettes. The result of a link approach will be identical in this Abelian case to the plaquette result, and only to be advocated for much larger Wilson loops. This applies for the Abelian case, while the essential advantage of the link approach comes when actually considering non-Abelian gauge theories.

\section{Conclusion and outlook}
\label{sec:conclusions}

Our objective in this work was to understand quantum simulation for non-local gauge-invariant quantities with time evolution. In particular, we have  successfully concentrated our efforts on the quantum simulation of space-time Wilson loops, for which we have presented a plaquette-based approach adequate for Abelian gauge models and a link-based approach applicable both to Abelian and non-Abelian gauge models. For the plaquette approach we have explicitly computed the time-oriented elementary plaquette in the temporal gauge for two models, and we have shown that the structure that appears in those two examples is general for models with a general electric plus magnetic (plaquette sum) Hamiltonian. Coming now to the link-based approach, we have introduced the two basic hopping operators, the quark/antiquark spatial line propagation hopping and the quark-antiquark creation/annihilation hopping Hamiltonian, out of which one can construct any space-time Wilson loop. The number of  ancillary degrees of freedom, additional to the pure gauge ones, is moderate, as they are reusable. The algorithm based on the link-approach is probabilistic in its success for non-Abelian models, deterministic for Abelian ones, and certain on success for both.

The space-time Wilson loop is relevant by itself and also in the context of transverse-momentum distributions, for instance, and the algorithm we propose here for its simulation can be implemented with current technologies for small sized Wilson loops. We have carried out a proof-of-principle numerical calculation for a small system size, that informs us as to the level of Trotterization likely to be required in a digital implementation.

This algorithm can potentially be applied to any light-front parton correlator, thus addressing one of the main obstacles of current lattice techniques, namely time dependence in parton correlators. In the current work we have  discussed only pure gauge models. This comes about because of the centrality of the Wilson line for any gauge invariant, non-local, space-time quantity. Thus, our proposal paves the way towards the quantum simulation of the generic situation. Indeed, for any realistic parton distribution matter fields are needed, and the next logical step in the development of the topic is the connection between our proposal for space-time dependent pure gauge objects with these matter fields. It should be emphasized though that the simulation of minimal coupling between the gauge and matter field has been the subject of many theoretical proposals~\cite{cloet2019opportunities,Banuls:2020yw} and there already exist scalable experimental realizations in  cold atom setups~\cite{Mil1128}.  While not trivial, a combination of those techniques with the algorithms presented here is certainly implementable in the foreseeable future, thus providing additional information and insight for hadron structure.

\section*{Acknowledgments}

We thank M. Engelhardt for valuable lattice-QCD insights during the early stages of the manuscript.

M.G.E. is supported by the Spanish Science and Innovation Ministry grant PID2019-106080GB-C21, and by the Community of Madrid and UAH joint grant CM/BG/2021-002 (MultiNuS) within the agreement to fund \emph{Beatriz Galindo} researchers.
E.R. thanks the QuantERA project QTFLAG. E.R. and I.L.E. acknowledge support of the Basque Government grant IT986-16.

\bibliographystyle{JHEP}
\providecommand{\href}[2]{#2}\begingroup\raggedright\endgroup

\end{document}